\documentclass{natureprintstyle}

\usepackage{graphicx}
\usepackage{amsmath,xcolor,url}
\usepackage{multirow}
\usepackage{longtable}
\usepackage{amssymb}
\usepackage{lineno} 
\renewcommand{\thefootnote}{\textit{\alph{footnote}}}

\usepackage{times}

\usepackage{graphicx}
\usepackage{amsmath}
\usepackage{amssymb}
\usepackage[final]{pdfpages}
\usepackage{epstopdf}
\usepackage{graphics,graphicx}
\usepackage{psfig}
\usepackage{threeparttable}

\usepackage{astjnlabbrev-nature} 

\title{Delayed Radio Flares from a Tidal Disruption Event}

\author{\large A. Horesh$^{1,\ast}$, S. B. Cenko$^{2,3}$, I. Arcavi$^{4,5}$\\
\\
\small{$^{1}$Racah Institute of Physics, The Hebrew University of Jerusalem, Jerusalem 91904, Israel}\\
\small{$^{2}$Astrophysics Science Division, NASA Goddard Space Flight Center, Mail Code 661, Greenbelt, MD 20771, USA}\\
\small{$^{3}$Joint Space-Science Institute, University of Maryland, College Park, MD 20742, USA}\\
\small{$^{4}$The School of Physics and Astronomy, Tel Aviv University, Tel Aviv 69978, Israel}\\
\small{$^{5}$CIFAR Azrieli Global Scholars program, CIFAR, Toronto, Canada}\\
\small{$^{*}$assafh@mail.huji.ac.il}\\
}

\usepackage[english]{babel}
\usepackage[utf8]{inputenc}
\usepackage{fancyhdr}

\newcommand{\ShortTitle}{The Radio Awakens - Delayed Radio Flares from a TDE}  

\begin{document}

\maketitle

\begin{abstract}

Radio observations of tidal disruption events (TDEs) - when a star is tidally disrupted by a supermassive black hole (SMBH) - provide a unique laboratory for studying outflows in the vicinity of SMBHs and their connection to accretion onto the SMBH. Radio emission has been detected in only a handful of TDEs so far. Here, we report the detection of delayed radio flares from an optically-discovered TDE. Our prompt radio observations of the TDE ASASSN-15oi showed no radio emission until the detection of a flare six months later, followed by a second and brighter flare, years later. We find that the standard scenario, in which an outflow is launched briefly after the stellar disruption, is unable to explain the combined temporal and spectral properties of the delayed flare. We suggest that the flare is due to the delayed ejection of an outflow, perhaps following a transition in accretion states. Our discovery motivates observations of TDEs at various timescales and highlights a need for new models.

\end{abstract}

Synoptic time-domain surveys have been increasingly fruitful in discovering nearby tidal disruption events (TDEs) over the last several years\cite{van_velzen_2020}. These transient events which are interpreted as
stars being tidally disrupted by super-massive black holes\cite{rees_1988} (SMBHs),
may provide a window into many diverse astrophysical
questions. Uncovering dormant SMBHs is only one of the
revelations made through these events. The complex physical process of
accretion of matter onto a SMBH is another. The 
nature of TDEs and their emission mechanisms are still puzzling. 
For example, what generates the ultraviolet (UV) and optical emission? Is it a process related to accretion or maybe internal shocks in streams of stellar debris\cite{piran_2015,jiang_2016}? Does an accretion disk form around the SMBH, and if so, when and in which geometry\cite{guillochon_2015,piran_2015,bonnerot_2016}? Panchromatic studies of a growing number of events hold the key for unlocking many of the remaining open
questions. 

An example is the progress made in recent years by uncovering
potentially two distinct sub classes of TDEs: (1) thermal TDEs,
discovered via their optical/UV emission, and (2) relativistic events such as {\it Swift }J1644+57 [ref.\citen{Levan_2011,Bloom_2011,Burrows_2011}], which exhibit high-energy non-thermal emission. Until recently, thermal TDEs were not seen to exhibit strong radio emission, but this changed with the discovery of a prompt radio signal\cite{Alexander_2016,van_Velzen_2016} from the nearby event ASASSN-14li [ref.\citen{Holoien_2016a}]. The weak radio emission has been interpreted as originating from a sub-relativistic shockwave launched into the SMBH circumnuclear material (CNM), driven by either outflows from an accretion disk\cite{Alexander_2016} or by the unbound stellar debris traveling away from the central SMBH\cite{Krolik_2016,Yalinewich_2019} (another plausible explanation is that the outflow is a jet that slowed down\cite{van_Velzen_2016}). On the other hand, {\it Swift}J1644+57 exhibited a strong radio afterglow\cite{Zauderer_2011,Berger_2012}, orders of magnitude more luminous than the emission in ASASSN-14li, originating
from a relativistic jet that is viewed on-axis. 

The field of TDE radio observations have seen additional developments in recent years. Very long baseline radio observations of an infrared (IR) transient that is considered a TDE candidate (Arp\,299B-AT1), revealed a radio jet\cite{mattila_2018}. Moreover, a recent radio transient, discovered independently of detection at other wavelengths, is also attributed to a TDE\cite{anderson_2019}. A recent excitement is the discovery of a coincident neutrino with an the optically discovered TDE\cite{stein_2020}, AT\,2019DSG, which also exhibits radio emission similar to the one in ASASSN-14li. These past discoveries are just additional pieces of the puzzle, which  will hopefully allow building a coherent picture of the overall physical processes in play.

In search for similar radio emission from other nearby TDEs,
we carried out radio ($6$ - $22$\,GHz) observations, using the Karl Jansky Very Large Array (VLA) telescope and recently also using the Arcminute Microkelvin Imager (AMI) telescope, of a number of optically discovered TDE candidates at
various time scales from early to late times (A. Horesh et al., manuscript in preparation). While most of the
observations resulted in null-detections, a single event, named
ASASSN-15oi, revealed a delayed radio flare, months after its optical discovery,  followed by somewhat even more surprisingly a second flare years later.

ASASSN-15oi was discovered in optical wavelengths by the All-Sky Automated Survey for
SuperNovae (ASASSN\cite{Kochanek_2017}) on 2015 August 14 [ref. \citen{Holoien_2016b}] at a distance of $216$\,Mpc. At the time of discovery
it had an optical magnitude of ${\rm V}\sim 16.2$ while previously it
was not detected on 2015 July 26 down to ${\rm V}\lesssim 17.2$, suggesting it was discovered relatively young. An optical spectrum obtained by the PESSTO
collaboration\cite{smartt_2015} on 2015 August 20, provided the initial
classification of ASASSN-15oi as a TDE\cite{Prentice_2015}. Upon discovery, multiple groups launched
panchromatic monitoring campaigns. In the optical, additional spectroscopy of ASASSN-15oi confirmed the initial classification of the event as
a TDE\cite{Holoien_2016b} and showed a rapid spectral evolution (compared to other optically discovered TDEs). A search for high-energy emission by the Neil Gehrels {\it Swift} Observatory, initially showed no significant emission\cite{Arcavi_2015}. However, deeper X-ray observations revealed what initially seemed as a non varying weak X-ray source, but which later increased in flux\cite{Gezari_2017,holoien_2018}. In the radio, we launched a monitoring campaign using the VLA. 

\section*{Discovery of a Delayed Radio Flare}

Our radio campaign began on 2015 August 22, eight days
after the optical discovery. Our initial observation performed in both $5$\,GHz and
$22$\,GHz, resulted in null-detections with $3\sigma$ limits of $\approx 33\mu$Jy ($1.8\times 10^{27} {\rm erg\,s}^{-1}\,{\rm Hz}^{-1}$) and $\approx 60\mu$Jy ($3.3\times 10^{27} {\rm erg\,s}^{-1}\,{\rm Hz}^{-1}$),
respectively. Despite the non-detections, we kept observing the
source, motivated by some theoretical models that suggest a delay between the TDE optical flare and the formation of the accretion disk\cite{shiokawa_2015,dai_2015,piran_2015}. We observed ASASSN-15oi
twice more on 2015 September 6 ($\Delta t=23$\,days, where $\Delta t$ is the time since optical discovery) and November 12 ($\Delta t=90$\,days), observations which also resulted in
null-detections, until the discovery of
significant radio emission on 2016 February 12 ($\Delta t=182$\,days; Figure 1 and Extended Data Figure 1), with an approximate flux density of $\approx 1300\,\mu$Jy ($7.3\times 10^{28} {\rm erg\,s}^{-1}\,{\rm Hz}^{-1}$), at a peak frequency of $\sim 9.6$\,GHz. Once ASASSN-15oi was detected in
the radio, we embarked on a followup observing campaign, carrying out observations in multiple radio frequencies in order to characterize the
properties of the broadband radio spectrum and its evolution.

Separately from our followup observing campaign, the position of ASASSN-15oi was observed recently as part of the Very Large Array Sky Survey (VLASS\cite{lacy_2020}). Inspection of the quick-look images (produced by the National Radio Astronomy Observatory [NRAO]) reveals a re-brightening of the radio emission at $3$\,GHz on 01  July 2019 (almost 4 years after the initial optical discovery), to a flux density level of $\approx 8000\,\mu$Jy ($4.4\times 10^{29} {\rm erg\,s}^{-1}\,{\rm Hz}^{-1}$). 

\section*{The Peculiar Evolution of the Delayed Radio Flare}

Our followup campaign (which includes six additional
observing epochs since radio discovery) reveal an unusually evolving radio spectrum
(see the full spectral evolution in Extended Data Figure~2). During the first $2$\,weeks after the initial detection of the radio emission, the
peak frequency slowly evolved, from $\nu_p =9.6\pm 0.9$ to $\nu_p=8.6\pm 0.4$\,GHz (see section `Radio Observations' in the Methods), while at the same time, the peak flux density dropped by $\approx 25\%$. In contrast, at later times, the radio flare evolved quickly. As seen in Figure~\ref{fig:peak}, the peak frequency of the radio emission decreased to $< 3$\,GHz only $2$ months after radio discovery.

Another oddity is the shape of the radio spectrum. In general, the radio spectral peak ($\nu_{p}$) observed in transient phenomena is usually due to either the minimum energy of the emitting electrons (in which case the flux density at $\nu < \nu_{p}$ is $F_{\nu} \propto \nu^{1/3}$) or due to absorption (usually due to synchrotron self absorption [SSA] and thus the flux density at $\nu < \nu_{p}$ is $F_{\nu} \propto \nu^{5/2}$, as seen for example in most radio supernovae). Neither of these two cases is observed in ASASSN-15oi. Instead, our data show that the flux density, $F_{\nu}\propto \nu ^{\alpha}$, has a power-law of $\alpha \approx 1$ (in contrast to both ASASSN-14li [ref.\citen{Alexander_2016}] and {\it Swift}J1644+57 [ref. \citen{Berger_2012}] in which $\alpha \approx 5/2$). Although this differs from the spectral shape of standard SSA (from a homogeneous CNM shockwave) model, it may be explained by a more complex model in which the SSA originates from an inhomogeneous source (see section `Spectral Modeling of the Radio Emission' in the Methods). This spectral index is also typical of radio emission originating from some quiescent GHz-peaked spectrum (GPS) sources\cite{odea_1990} which are considered to be young active galactic nuclei (AGNs), but do not exhibit flares on timescales as observed here (see section `Spectral Modeling of the Radio Emission' in the Methods and Extended Data Figure~3). The spectral index of the optically thin synchrotron emission above the peak ($F_{\nu>\nu_{p}}\propto \nu^{\alpha_{syn}}$), is a function of the power-law index ($p$) of the energy distribution of the emitting electrons ($N_{e} \propto E^{-p}$). For ASASSN-15oi, the spectral index is initially in the range $-2\leq\alpha_{syn}\leq-1$, which is somewhat steeper than the spectral index observed in both {\it Swift}J1644+57 and ASASSN-14li. Later, when the peak flux density decreased below $3$\,GHz, the optically thin spectrum became shallower ($\alpha_{syn}\gtrsim -0.6$; which is also shallow compared to ASASSN-14li and {\it Swift}J1644+57).

\section*{The Nature of the Delayed Radio Flare}

The delayed radio emission we observe from the TDE ASASSN-15oi, its properties and evolution raise several key questions. Can the late-time radio emission and the earlier null-detections be reconciled under a standard CNM shockwave model? If so, does it originate in a relativistic jet, such as observed in {\it Swift}J1644+57 or possibly an off-axis jet which becomes visible only at late times? Or does it point to a sub-relativistic shockwave (driven by accretion disk outflows or stellar debris) such as the one observed in ASASSN-14li? Does a delayed radio detection require a delayed outflow formation? 

To address these questions, first we test the standard dynamical models that have been used to explain the radio emission from TDEs so far\cite{giannios_2011,metzger_2012,chevalier_1998,generozov_2017}. In these models, a single shockwave (either relativistic or sub-relativistic) is launched into the CNM around the time of optical discovery. Both the optically thick and optically thin emission have a power-law temporal evolution with a range of values for the power-law indexes (depending on the properties of the shockwave; see section `Temporal Evolution of the Radio Emission' in the Methods). The steep rise of the observed radio emission from non-detection on $\Delta t=90$\,days to detection on $\Delta t=182$\,days requires that the temporal evolution of the flux density will be steeper than $F_{\nu}\propto t^{4}$. 
In the analytical models that we explore, the fastest rise rate of the emission occurs in the relativistic jet case. 
When an on-axis relativistic jet is interacting with CNM profiles in the range between a constant density to $\rho_{\rm CNM} \propto r^{-2.5}$ (which is the steepest density profile found so far in TDEs\cite{alexander_2020}), the flux density fastest rise is $F_{\nu}\propto t^{3}$, when the emission is optically thick (for comparison, the optically thick radio emission in a sub-relativistic supernova usually results in shallower rise of $F_{\nu}\propto t^{2.5}$). Extrapolating backward in time from our initial radio detection in C-band on 2016 Feb 12, results in a predicted flux density of $0.15$\,mJy on 2015 Nov 12, well above our detection limit (see section `Temporal Evolution of the Radio Emission' in the Methods). A steeper rise can be obtained if the relativistic jet is observed off-axis. Exploring numerical models of such a scenario\cite{van_eerten_2012}, we find that it cannot account for both the steep rise from non-detection to detection, and for the subsequent spectral and temporal evolution of the detected radio emission (see section `Temporal Evolution of the Radio Emission' in the Methods). Another possible model, used to explain the re-brightening of radio emission at late times (such as the one observed in the relativistic TDE {\it Swift}J1644+57 [ref. \citen{Berger_2012}]) is a structured relativistic jet\cite{Mimica_2015,granot_2014} launched promptly after stellar disruption. In this case, the slower parts of the jet, will provide additional power source for the observed emission in late times. However, even in this case the steepest rise due to such a jet structure will result at most in a $\approx t^{3}$ increase in flux density which is not enough to explain the steep rise in the initial radio flare we observe. 
The big jump in flux density that we observe from null-detection to detection thus seems to disfavor the predictions of the existing models that we discussed above, that invoke the interaction of a relativistic (on or off-axis) or sub-relativistic outflow with the CNM promptly after the disruption of the star, and therefore points to a radio emitting process that occurs at late-times. 

The temporal evolution of the optically-thin emission, following its temporal peak, is inconsistent with the above models. Whether we consider an on- or off-axis relativistic jet or a sub-relativistic outflow, the optically thin radio emission is expected to have a power-law temporal evolution ($F_{\nu} \propto t^{\beta}$) with a predicted\cite{Krolik_2016,generozov_2017} power-law index of $-1 \geq\beta\geq -3$. However, here, we see a varying rather steep temporal evolution, where the temporal power-law reaches a value $\beta < -3$ (see Extended Data Figure~4).

A possible solution for explaining the initial radio null-detections is if unbound material from the TDE was initially traveling in a cavity around the SMBH until it suddenly reached an extended high-density CNM structure. 
However, once the outflow enters the high-density extended CNM structure, the radio emission from the shockwave should follow a spectral evolution similar to ASASSN-14li (or {\it Swift}J1644+57) in contrast to what we observe (see Figure 2 and Extended Data Figure~4) making this scenario less plausible. A scenario in which an outflow from the TDE suddenly encounters a spatially thin confined dense CNM shell or filament, producing a brief late-time flare, is also disfavored. In this latter case, the radio flux density is expected to decrease extremely rapidly after peak ($F_{\nu}\propto t^{-9}$), once the outflow crosses the filament\cite{harris_2016}, but such a steep decline is not observed here. One may also consider the possibility that an outflow was launched only at late-times due to delayed accretion of bound stellar debris (as suggested by some studies\cite{shiokawa_2015,piran_2015}). But also in this case, once the outflow is launched into the CNM, the resulting radio emission should have spectral and temporal properties according to the standard model discussed above, in contrast to the data presented here. It is possible though that some delayed activity (accretion) is in play, but one that involves processes that are currently not included in standard TDE models.

The re-brightening of the radio emission on July 2019 may be the most recent evidence against the scenarios we examine above. A secondary flare with an increase in emission by a factor of $> 20$, years after the onset of the TDE, is not expected in any of the above scenarios. Even if considering a radio re-brightening in a structured jet model\cite{Mimica_2015,granot_2014}, then currently, none of these models predict a re-brightening of the flux density by more than an order of magnitude over such a long time scale as observed here. Moreover, explaining a re-brightening at late times with a structured jet that has been launched early on, requires that the initially observed delayed late-time emission will be explained by this outflow as well. However, as we already showed, it cannot. A secondary flare, years after the onset of the TDE, is therefore not expected in any of the above scenarios. One possibility is that the re-brightening is driven by the same process responsible for the initial delayed flare we detected. One may also consider the possibility of a recurring TDE flare due to repeated partial disruptions of a star\cite{gul_2013,campana_2015}. Another possible explanation is that the TDE occurred around a binary SMBH system. In such a scenario, the accretion rate may be highly variable with multiple peaks including also several years after the initial disruption\cite{coughlin_2017}. Still, all of these proposed theoretical explanations currently do not offer clear predictions for late-time radio emission that can be tested against our measurements. Furthermore, unfortunately, information about the 2019 re-brightening event is also limited (K. Alexander et al., manuscript in preparation).

Late-time ultraviolet (UV) and optical observations show no signature of any renewed activity or a secondary flare during the first year after the optical discovery of ASASSN-15oi [ref. \citen{holoien_2018}]. UV emission is still detected from the TDE after the time in which we discovered the delayed radio flare (with observations ongoing up to a year after optical discovery), and is consistent with a simple power-law decline of the UV emission detected at early times. A series of optical spectra taken starting at $\Delta t = 301$\,days and up to $\Delta t = 455$\,days shows that the broad emission lines, typical of TDEs, which are detected early on, have diminished, and no new emission lines are present\cite{holoien_2018}.  

We now compare the evolution of the X-ray emission with that of the radio emission (Figure 3). The X-ray emission, that is detected early on slowly and steadily rose with time, and peaked (after an increase in flux by a factor of $\sim 10$) about a year after the optical discovery\cite{Gezari_2017, holoien_2018}. A direct comparison with the evolution of the radio emission is somewhat limited by a gap in X-ray data during the time of the first late-time radio detection and the radio subsequent followup observations. It is clear that the radio emission does not increase in parallel to the X-ray emission (Figure 3), as the radio emission already fades away at $\lesssim 200$\,days, while the X-ray emission is still rising. Interestingly enough, the X-ray emission, after the gap in X-ray observations increased beyond $1\%$ of the Eddington luminosity, and its thermal (soft) component became brighter\cite{holoien_2018}.This behaviour is usually observed in X-ray binaries (XRBs), although while this transition is observed at a level of $\sim 1\%$ in some XRBs, it usually occurs on average at higher Eddington luminosities ($\sim 10 - 30\%$) [\cite{Dunn_2010, Maccarone_2003, Tetarenko_2016}]. A possible explanation of this behaviour in XRBs is that this transition occurs when the accretion rate increases and fresh material with a high Lorentz factor is injected into an existing jet\cite{Fender_2004}. Following this stage in XRBs, a radio flare is observed\cite{Fender_2004}. 
We also note that combined X-ray observations during June - August 2019, the period in which a re-brightening of the radio emission is detected in VLASS, show that the X-ray flux slightly increased, after declining, to a flux level of $8.07\pm1.1 \times 10^{-14}{\rm erg\,cm}^{-2}\,{\rm s}^{-1}$. This translates to $0.4\%$ of the Eddington luminosity only (lower than the X-ray luminosity increase to a level of $1\% L_{Edd}$ around the time of the initial delayed radio flare we observed). However, these recent X-ray data are limited and averaged over several months (see section `A Comparison Between the X-ray and Radio Emission Temporal Evolution' in the Methods), thus making their interpretation difficult.

Radio emission with a similar spectral shape as the one we observe in ASASSN-15oi and similar temporal behaviour (in contrast to the one in GPS sources), has been observed in some AGN or blazar radio flares (but which is not necessarily typical of the whole AGN/blazar flare population).
In September 2011, a radio flare from M81 exhibited an inverted radio spectrum with a peak frequency of $\sim 10$\,GHz. The flare radio flux density slowly decreased on a timescale of weeks with the spectral peak frequency roughly staying the same, until a second flare was observed\cite{king_2016}. In another case, a year long radio flare from the blazar CTA\,102 had a complex temporal and spectral behaviour\cite{fromm_2011}. The late time evolution phase of this radio flare also shows similar characteristics to those observed in ASASSN-15oi (Figure 2). In general, these AGN/blazar radio flares have been partially explained by the shock-in-jet model\cite{marscher_1985} in which a shock propagates in an existing radio jet leading to what seems as a flare. However, a phase in which the radio peak frequency does not vary while the peak flux density decreases, as observed in both the M81 and the CTA\,102 flares, is not captured by this model. Nonetheless, it is possible that there is a radio weak quiescent jet associated with the SMBH of ASASSN-15oi, which is shocked by a delayed injection of energy by the TDE. Such preexisting weak quiescent radio emission, that suggests a non-TDE related activity of the SMBH, has been found in ASASSN-14li\cite{Alexander_2016}, at a level which is too faint for detection at the distance of ASASSN-15oi.

It has been suggested, that emission in both XRBs and AGNs is dominated by a weak non-thermal jet when the accretion rate is considerably sub-Eddington (the accretion becomes radiatively inefficient)\cite{Falcke_2004}. A phase transition in accretion occurs when the accretion rate increases above a critical threshold, at which point the emission in X-ray becomes disk dominated and a high velocity outflow is launched (sometimes observed as spatially discrete knot ejections\cite{bright_2020, king_2016}) resulting in a radio flare. It is therefore possible that such a phase transition occurred in ASASSN-15oi resulting in the delayed launch of an outflow which led to rapidly-rising radio emission at late times. However, what triggers this phase transition, how this transition and the outflow launching coupled to it depend on the nature and properties of the relevant phenomena (e.g., TDE vs XRB), and what the typical signatures of this transition in TDEs are (e.g will all TDEs exhibit an increase of their X-ray emission above a certain threshold characterized by some percentage of their Eddington luminosity), are still open questions. The details of what follows any transition in the accretion phase are also unclear. Recall, that once detected, we observed an initially slowly evolving inverted spectrum, but, shortly after, the spectral peak frequency rapidly evolved. It is possible that the termination of the slow spectral evolution phase marks the point at which the shock reached the edge of a pre-existing jet and that the emission that follows is of the slowly cooling jet. Testing this scenario and answering any other open question will hopefully become possible with the discovery of additional events such as ASASSN-15oi. 

Yet another open question relates to how common are such late-time flares (due to a delayed outflow ejection) in TDEs. There is one other case of a possible late-time radio flare in the IR TDE ARP\,299B-AT1 [ref. \citen{mattila_2018}]. In that case, a single frequency (8.4\,GHz) radio observation, taken 12 days after the first possible indication of an increase in the IR flux, resulted in a null-detection. The next observation, carried out just $48$\,days later, detected  increasingly bright radio emission. The full set of radio measurements (spanning thousands of days) of  ARP\,299B-AT1 is consistent with a relativistic jet launched briefly after the stellar disruption. The late-time radio spectrum, which is consistent with originating from an electron energy distribution of $N_{e} \propto E^{-3}$, was also slowly evolving, in agreement with the predictions of known models\cite{mattila_2018}, but in contrast to the evolution observed in ASASSN-15oi. 

It is possible that the delayed launch of an outflow, as we observe here, has been missed in other TDEs due to limited observational coverage. First, in several past TDEs there is a substantial gap between the time of disruption and the time at which the first radio observation is carried out (e.g. the disruption time in the case of ASASSN-14li is poorly constrained). Thus even if radio emission is detected initially in such cases, the exact time at which the outflow was launched with respect to disruption is unknown. Moreover, radio observations, in most cases, whether radio emission is detected or not, are curtailed after several months, leaving any flaring event occurring later on, undetected. The peculiar delayed flares we discovered in ASASSN-15oi on time scales of months and years thus motivate carefully planned observational campaigns of TDEs from early times until very late times.

\section*{Summary}

Our radio observing campaign of the optically discovered TDE ASASSN-15oi since discovery and up to more than a year later, revealed a delayed radio flare with odd spectral and temporal properties. A second even more luminous radio flare has been detected in VLASS observations. The various models that we explore here, that have been suggested so far for explaining the radio emission originating from TDEs, are unable to explain the combined properties of the observed radio emission. Specifically, it seems that such a delayed bright radio flare, following an extended period of null radio detections, requires some sort of an outflow to be launched at late times (into a possibly inhomogeneous CNM; see Methods), suggesting a delayed onset of enhanced accretion. Some of the properties of the emission have similarities to XRBs and to AGN/blazar radio flares, thus raising the possibility that a transition in accretion phase state, that is proposed as an explanation of these latter flares, is also in play in TDEs. The details of this process, that has not been observed in TDEs until now, and what triggers it are yet to be discovered. 
Understanding this process, requires that we first better characterize it. Our discovery thus motivates late-time radio campaigns of TDEs that will hopefully identify additional delayed flares. These could help us study the process responsible for triggering delayed enhanced accretion, the subsequent outflow launching, and the emission that accompanies it, thus helping in unveiling the nature of this new puzzling phenomenon in TDEs.


\bibliography{ASASSN15oi.bib} 



\begin{addendum}
\item[Acknowledgements] 
We thank the anonymous referees for improving this manuscript, and T. Piran and E. Nakar for useful discussions.  A.H. is grateful for the support by grants from the I-CORE Program of the Planning and Budgeting Committee and the Israel Science Foundation (ISF), and from the US-Israel Binational Science Foundation (BSF). 
I.A. is a CIFAR Azrieli Global Scholar in the Gravity and the Extreme Universe Program and
acknowledges support from that program, from the European Research Council (ERC)
under the European Union’s Horizon 2020 research and innovation program (grant
agreement number 852097), from the Israel Science Foundation (grant number 2752/19),
from the United States - Israel Binational Science Foundation (BSF), and from the Israeli
Council for Higher Education Alon Fellowship.
The authors thank the NRAO staff for approving and scheduling the VLA observations. The National Radio Astronomy Observatory is a facility of the National Science Foundation operated under cooperative agreement by Associated Universities, Inc. The authors thank the {\it Swift} TOO team. This research has made use of data and/or software provided by the High Energy
Astrophysics Science Archive Research Center (HEASARC), which is a service of the
Astrophysics Science Division at NASA/GSFC. This research has made use of the CIRADA
cutout service at URL cutouts.cirada.ca, operated by the Canadian Initiative for Radio
Astronomy Data Analysis (CIRADA). CIRADA is funded by a grant from the Canada
Foundation for Innovation 2017 Innovation Fund (Project 35999), as well as by the Provinces of Ontario, British Columbia, Alberta, Manitoba and Quebec, in collaboration with
the National Research Council of Canada, the US National Radio Astronomy Observatory
and Australia’s Commonwealth Scientific and Industrial Research Organisation.

\item[Author contributions] 

A.H. has led the radio observing campaign, the data analysis and modeling, the interpretation, and the manuscript preparation. S.B.C and I.A. contributed to the interpretation of the results  and to the manuscript preparation.

\item[Correspondence and requests for materials] should be addressed to A.H. (email: assafh@mail.huji.ac.il).

\item[Additional information]

Extended Data Figures and Supplementary information are available for this paper.


\item[Competing interests]
 The authors declare that they have no competing financial interests.

\item[Data availability]

The ASASSN-15oi radio data, presented in several plots, can be found in Supplementary Table~1. The raw VLA data is available on the NRAO archive at \url{https://archive.nrao.edu/archive/advquery.jsp}. Tools to analyze the VLA data can be found at the NROA website at \url{https://science.nrao.edu/facilities/vla/data-processing/analysis-packages/analysis-packages}. The collection of radio data of other TDEs can be found in [ref. \citen{alexander_2020}]. The ASASSN-15oi x-ray emission measurements can be found in [ref. \citen{Gezari_2017}]. Any additional data that support the findings of this study are available from the corresponding author upon reasonable request.

\end{addendum}


\clearpage

\begin{figure*}
    \centering
    \includegraphics[width=1\linewidth]{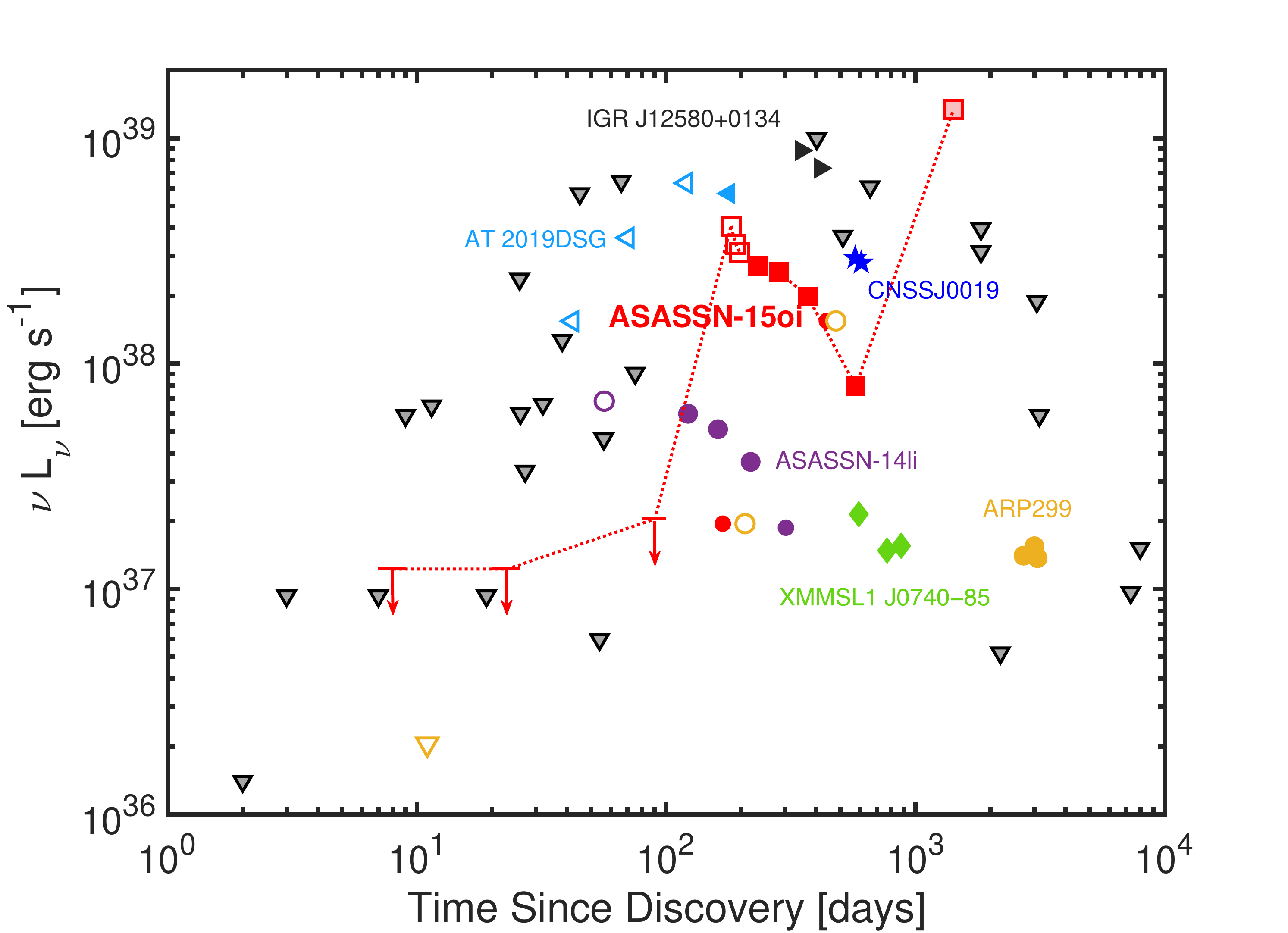}
    \caption{\textbf{The radio luminosity of a handful of TDEs (including the delayed radio flare from ASASSN-15oi) as a function of time.} Note, that the time here is after optical discovery (in some cases the actual disruption time, such as in ASASSN-14li, is poorly constrained, thus the light curves of these events may shift in time compared to others). The radio measurements (luminosity densities $L_{\nu}$) are at a frequency of $\sim 5$\,GHz (data for AT\,2019dsg is at $8.5$\,GHz). Empty markers represent a phase where the emission is optically thick while filled markers represent optically thin emission. The radio emission of ASASSN-14li (purple circles\cite{Alexander_2016}) and AT\,2019DSG (light blue side triangles\cite{stein_2020}) shows a typical evolution of a shockwave in a CNM as the emission peaks when the optical depth is about unity at the relevant frequency and thereafter the emission declines (similar to what is also seen in radio supernovae). The radio emission from the radio discovered TDE, CNSS\,J0019 (ref. \citen{anderson_2019}), exhibits similar behaviour as in the two latter TDEs (the radio peak is not shown here as it is at lower GHz
frequencies). The radio observations of XMMSL1 J0740-85 (green diamonds\cite{Alexander_2017}) and IGR J12580+0134 (blue stars\cite{irwin_2015}) took place only at later times, when the radio emission is already optically thin and fading. The radio available for ARP\,299B-AT1 (orange circles\cite{mattila_2018}) suggests the radio emission originated from a relativistic jet launched promptly after disruption. In contrast, our radio observations, early after discovery and up to $90$\,days, resulted in null-detections ($3\sigma$ limits are represented by red line arrows). The delayed radio flare we detect later (red squares) has a peculiar evolution. The light red square at $\Delta t \sim 1400$\,days is the recent detection of a rebrightening emission from ASASSN-15oi at $3$\,GHz (lacking spectral information).  We also present past events with no detected radio emission (detection limits shown as gray triangles). The emission from {\it Swift}J1644+57 (not shown) is a few orders of magnitudes brighter than the emission detected from the TDEs presented in the figure, and slowly rising in the first $100$\,days after discovery.}
    \label{fig:TDE_lc_compare}
\end{figure*}

\clearpage

\begin{figure*}
    \centering
    \includegraphics[width=1\linewidth]{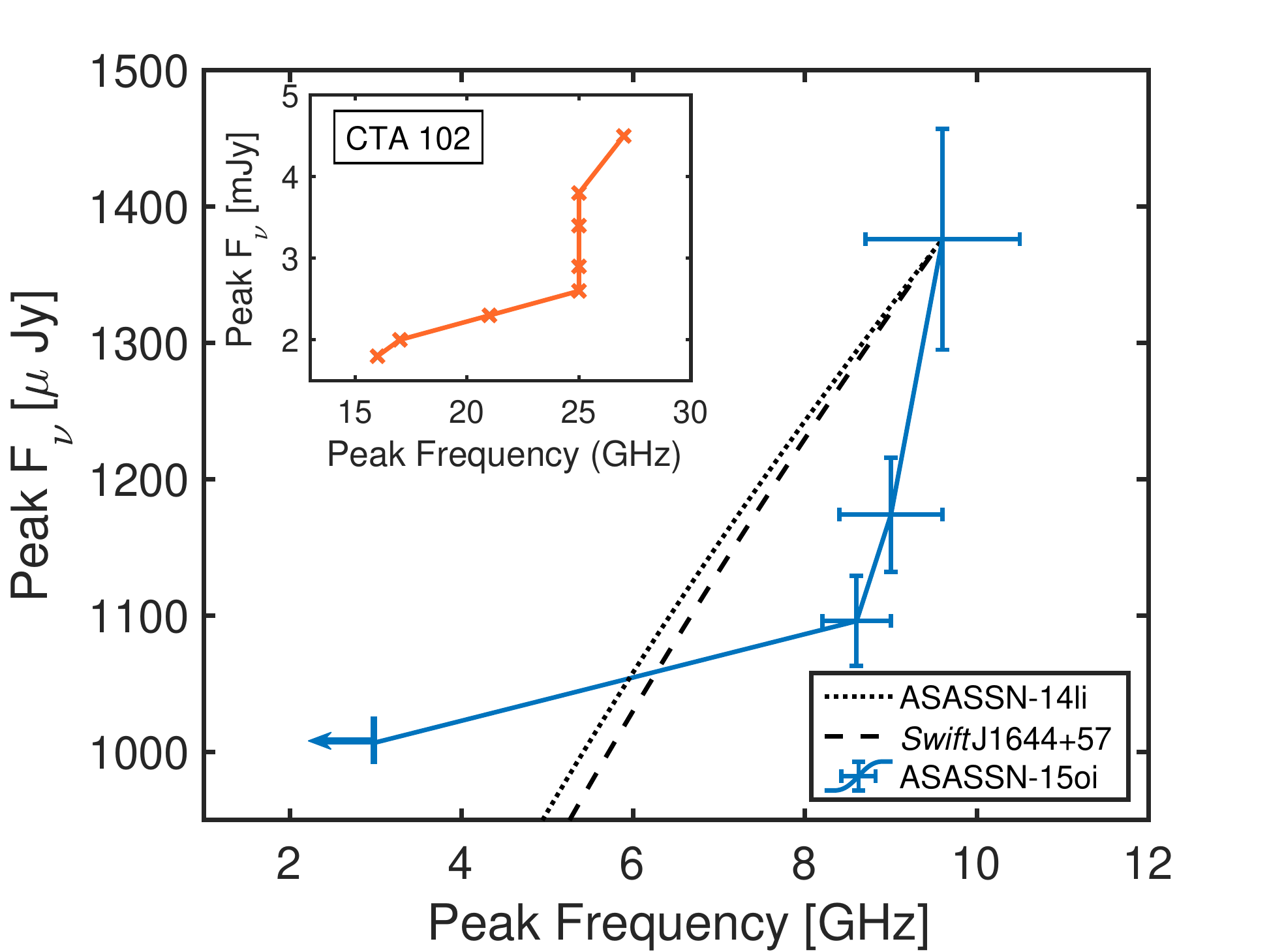}
    \caption{\textbf{The evolution of the peak flux density and frequency of the delayed radio emission from ASASSN-15oi.} During the first two weeks of the radio discovery the peak frequency evolves slowly, while in the six weeks that followed, the peak frequency evolves rapidly to an unknown frequency below $3$\,GHz (blue line and markers; peak limit marked by blue arrow; $1\sigma$ uncertainties from the peak best fit analysis are presented). In comparison, we show the evolution of the peak in both ASASSN-14li [ref. \citen{Alexander_2016}] (dotted line) and {\it Swift}J1644+57 [ref.\citen{Zauderer_2011}] (dashed line). This type of diagnostic plot of the radio peak is used sometimes to follow the evolution of radio flares from AGNs/Blazars. The peak behaviour of ASASSN-15oi is similar to the observed late-stage evolution of two AGN/blazar flares (see main text). The inset shows a late-time phase in the evolution of the radio flare from the blazar CTA\,102 [ref.\citen{fromm_2011}]. We emphasize, however, that this behaviour is not observed in all AGN/blazar flares.}
    \label{fig:peak}
\end{figure*}

\clearpage

\newpage

\begin{figure*}
    \centering
        \includegraphics[width=1\linewidth]{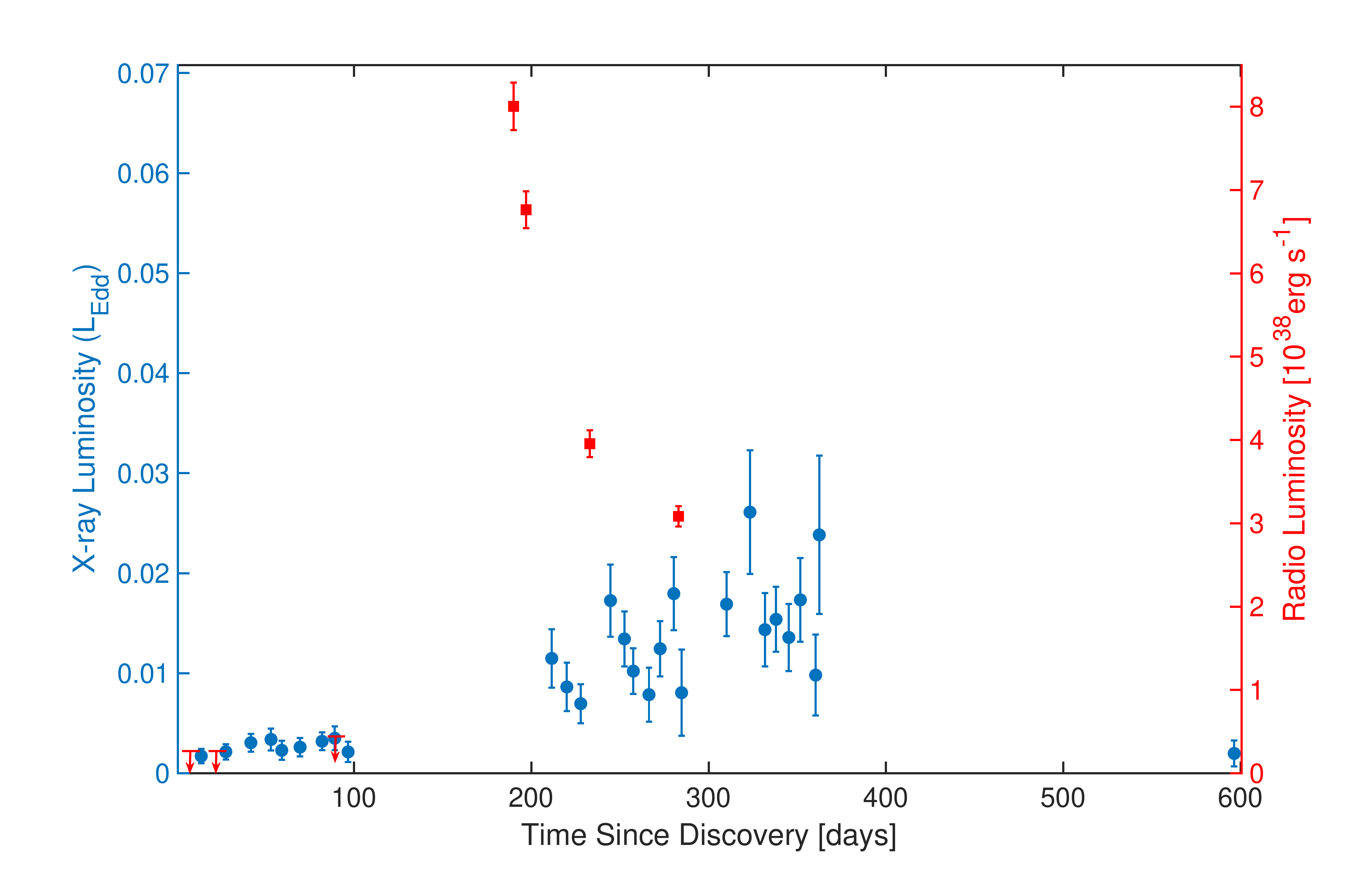}

    \caption{\textbf{Comparison of the temporal evolution of the X-ray luminosity with the optically thin radio luminosity in ASASSN-15oi.} The X-ray emission is detected soon after the occurrence of the TDE, while the radio emission begins later. Unfortunately, no X-ray data are available when the radio emission is initially detected. The x-ray luminosity, however, increases to a level of $\approx 1\%$ the Eddington luminosity when radio emission is detected. The X-ray data are presented as blue circles\cite{Gezari_2017} ($1\sigma$ errors are presented). The radio data is at a frequency of $13$\,GHz (red squares; uncertainties are the image noise and flux calibration uncertainty added in quadrature as defined in Supplementary Table~1). Averaged early radio 3$\sigma$ non-detections up to $90$ days after optical discovery are represented with red arrows.}
    
    \label{fig:xray}
\end{figure*}

\newpage



\clearpage

\section*{Methods}
\smallskip
\smallskip

\makeatletter 
\renewcommand{\thesection}{S\@arabic\c@section}
\makeatother

\makeatletter 
\renewcommand{\figurename}{Extended Data Figure}
\makeatother
\makeatletter 
\renewcommand{\tablename}{Supplementary Table}
\makeatother

\setcounter{section}{0}
\setcounter{figure}{0}
\setcounter{table}{0}

\section*{Radio Observations}
\label{sec:radio_obs}

We observed the field of ASASSN-15oi with the VLA on 2015 August 22, September 06, November 12, and on 2016 February 12, under a {\it Swift}-VLA joint program (SB\,4220). Later observations were performed under a director discretionary time (DDT) program (16A-422). The initial four observations were performed in X- and K-bands ($6$\,GHz and $22$\,GHz, respectively) only, as a detection experiment in search for radio emission. Once radio emission was detected in the fourth observation (Supplementary Figure~S1), the followup observations were conducted in a varying wide range of bands from S-band ($3$\,GHz) to K-band, as needed, in order to characterize and capture the evolution of the broadband radio spectrum (see Supplementary Figure~S2).

 We calibrated the radio data using the automated VLA calibration pipeline available in the Common Astronomy Software Applications (CASA) package\cite{2007ASPC..376..127M}. Flux density calibration was conducted using 3C48, while J2040-2507 was used as a gain calibrator. Images of the ASASSN-15oi field were produced using the CASA task CLEAN. In images where ASASSN-15oi was detected, the source flux density was measured using the CASA task IMFIT, and the image rms was calculated using the CASA task IMSTAT. In addition, we add a flux density calibration error at a conservative level of $3\%$ and $5\%$ to frequencies below and at (or above) Ku-band, respectively. The log of the observations and the resulting measurements are listed in Supplementary Table~1. 
 
 As mentioned in the main text, a re-brightening of the delayed flare is detected in VLASS data. VLASS data has been obtained over a long period of time and will continue to be collected in the following years. The observation in which ASASSN-15oi was detected was obtained on 2019 July. The data were reduced and imaged by NRAO using a software pipeline designed for VLASS. The flux density from ASASSN-15oi in the VLASS quick-look image is at a level of $\approx 8$\,mJy (with an assumed general uncertainty level of $15\%$ for all quick look images\footnote{https://science.nrao.edu/vlass/users-guide-to-quick-look-images}).

\section*{Spectral Modeling of the Radio Emission}\label{sec:spectral_modeling}

Below we attempt to model the observed broadband radio spectra of ASASSN-15oi at the individual observing epochs where a peak in the flux density is observed, but without modeling the temporal evolution. Theoretical models\cite{giannios_2011,metzger_2012,chevalier_1998,generozov_2017, Krolik_2016} predict that radio emission in TDEs originates from a forward shockwave (either relativistic or sub-relativistic) traveling in the surrounding environment. This shockwave accelerates free electrons that gyrate in the shockwave enhanced magnetic field and thus emit synchrotron radiation. Therefore, we first model the individual single-epoch broadband radio spectra, we observe, according to a synchrotron self absorption (SSA) spectral emission model\cite{chevalier_1998,weiler_2002,Barniol_2013}. This model successfully accounts for the radio emission observed in both {\it Swift}J1644+57 and ASASSN-14li [ref.\citen{Zauderer_2011,Berger_2012,Alexander_2016,Krolik_2016}].

In the SSA emission model, the radio spectrum exhibits a peak below which the emission is self absorbed and thus optically thick and above which the emission is optically thin. The optically thick emission can be described as  
\begin{equation}
    F_\nu \propto \frac{\pi R^2}{D^2} B^{-1/2} \nu^{5/2}, 
    \label{eq: optically thick}
\end{equation}
while the optically thin emission is described by 
\begin{equation}
    F_\nu \propto \frac{4 \pi f R^3}{3 D^2} N_0 B^{(p+1)/2} \nu^{-(p-1)/2}, 
    \label{eq: optically thin}
\end{equation}
where $R$ is the radius of the radio emitting shell, $D$ is the distance to the TDE, $f$ is the emission filling factor, and  $B$ is the magnetic field strength. The energy density of the magnetic field is a fraction $\epsilon_{\rm B}$ of the energy density of the shocked CNM. Thus, the magnetic field strength depends also on the square-root of the CNM density (and its profile). 

Our SSA best fit models of each of the ASASSN-15oi radio spectra, separately, in which a radio peak is evident, are presented in Supplementary Figure~S3. As shown in the figure, the SSA models poorly account for the observed radio spectra (with reduced $\chi^{2}$ values of $\chi_{r}^{2} > 8$ in times $\Delta t = 190, 197$\, days). This is no surprise, as the optically thick spectral index of the SSA model is $\alpha = 5/2$, while examination of the data suggests a spectral index of $\alpha \approx 1$. A shallower spectral index of the optically thick emission is expected if only internal free-free absorption (FFA) is the dominant absorption mechanism instead of SSA, although still steeper than the observed $\alpha \approx 1$ spectral index. 
In this internal FFA model\cite{weiler_2002} the flux density is: 
\begin{equation}
    F_\nu \propto F_{\nu,syn} \left(\frac{1-e^{-\tau_{ff}}}{\tau_{ff}}\right), 
\end{equation}
where $F_{\nu,syn}$ is the un-absorbed synchrotron emission, and $\tau_{ff}$ is the FFA optical depth. The results of the FFA modeling are presented in Supplementary Figure~S3. While the internal FFA models are somewhat better in accounting for the observed radio spectra than the SSA model ($\chi_{r}^{2} \approx 2.5$), they still significantly deviate from the observations. A solution may be found by reverting to the SSA model, but this time instead of assuming a homogeneous CNM environment, we will assume an inhomogeneous one. Inhomogeneities in the CNM can be modeled as inhomogeneities of the magnetic field and will result in a SSA spectrum with a broader peak and a shallower spectral index\cite{bjornsson_2017}. This explanation was also used recently to describe a shallow optically thick radio emission in a stripped envelope supernova\cite{chandra_2019}. We follow this model (which is an extension of the SSA model but one that is parameterized with a distribution of magnetic fields, $P(B) \propto B^{-a}$, instead of a single magnetic field). This model thus adds two additional degrees of freedom: the range of magnetic field strengths and the power-law index ($a$) of the magnetic field distribution. The best fit results of this model are presented in Supplementary Figure~S3. We find that the inhomogeneous SSA model provide a better spectral fit ($\chi_{r}^{2} \approx 0.8$) compared to the previous models. It is important to note that finding a separate good spectral fit to each of the individual observed spectra, does not mean that we have found a dynamical single scenario that can explain the combined full observed dataset, as we explain below when attempting a temporal modeling.

At this point, it is worth mentioning that the observed radio spectrum of ASASSN-15oi is reminiscent of GPS sources. These radio sources have a spectrum with peak frequency in the low GHz range, as their name suggests. They are powerful compact ($\lesssim 1$\,kpc) radio sources some of which exhibit a morphology of two sided symmetric sources, when resolved with high angular resolution observations, and are believed to be young AGN radio jets\cite{odea_1998}. On a long time scale, GPS sources can exhibit strong variability up to an order of a magnitude in the radio. However, the complex spectral variations observed here, over a short timescale of two months, are atypical of GPS sources. Still, considering that the radio spectral shape of GPS sources is attributed to inhomogeneities and the resemblance of their spectra to that of ASASSN-15oi strengthen the conclusion that the radio emission from ASASSN-15oi may originate from a complex CNM environment.

Despite the poor fit by a simple SSA model, assuming that the peak flux density is the result of SSA, we use the peak flux density and frequency to roughly estimate the shockwave radius\cite{chevalier_fransson_2006} by

\begin{align}
   \nonumber R_p  &= 4.0 \times 10^{14}  \left(\frac{\epsilon_{e}}{\epsilon_{B}}\right)^{-1/19} \left( \frac{f}{0.5} \right)^{-1/19} \left( \frac{F_p}{\rm{mJy}} \right)^{9/19} \left( \frac{D}{\rm{Mpc}} \right)^{18/19} \\ 
   &\times \left( \frac{\nu_p}{5\,\rm{GHz}} \right)^{-1}  \rm{cm}, 
\end{align}
where $\epsilon_{e}$ and $\epsilon_{B}$ are the fractions of shockwave energy deposited into accelerating free electrons and enhancing the magnetic field, respectively. We adopt standard equipartition value of $\epsilon_{e}=\epsilon_{B}=0.1$ and $f=0.5$.

The radius of the radio emitting region, when we first detect it (2016 February 12, $\Delta t=182$\,days), is estimated at $R\approx 4\times 10^{16}$\,cm, which implies (assuming an outflow was launched at optical discovery) a shockwave velocity of $25,000$\,km/s (only slightly higher than the velocity expected for the unbound stellar debris\cite{Krolik_2016}). Note, however, that if a jet (or an outflow) is launched later, then the velocity estimate will be higher. We also find that the radio-emitting region's radius remains roughly the same, over a two weeks period (when evaluated on 2016 February 20 and 27; $\Delta t=190$ and 197\,days, respectively, following the initial radius evaluation on $\Delta t=182$\,days). Following this period, the estimated radius makes a big jump to $R \gtrsim 1.1\times 10^{17}$\,cm, in only six weeks. The above radius estimates (as well as the velocity estimates), however, become lower limits if the  emission is originating from an inhomogeneous source (as discussed above). 

\section*{Temporal Evolution of the Radio Emission}\label{sec:temporal_modeling}

The temporal evolution of the observed radio emission can be examined against main theoretical predictions. We first address the analytical predictions for an on-axis relativistic jet and a sub-relativistic outflow. In the case of an on-axis relativistic jet\cite{metzger_2012} the optically thick emission is expected to rise as $t^{(k+2)/(4-k)}$, where $k$ is the power-law index of the CNM density profile ($\rho_{\rm CNM}\propto r^{-k}$). Adopting the steepest density profiles found in some TDEs\cite{alexander_2020} of $\rho_{\rm CNM}\propto r^{-2.5}$ (which is steeper than the usual wind-like CNM profile used in most numerical simulations), we obtain $F_{\nu} \propto t^{3}$. In the sub relativistic case\cite{Krolik_2016}, the optically thick emission evolves as $F_{\nu,thick}\propto t^{(28+3.5k)/14}$
, which is shallower than the relativistic case. We also found that attempting to model the full observed data, including the observed non-detections, with a sub-relativistic spherical outflow model, fails. 
Adopting, therefore, the steepest relation $F_{\nu} \propto t^{3}$ (of the relativistic case), the optically thick emission detected on 2016 Feb 12 can be evolved back in time to 2015 Nov 12, resulting in a predicted flux density level of $0.15$\,mJy in C-band, well above the detection threshold of our observation at that time (a $3\sigma$ limit of $60\,\mu$Jy). As seen in Supplementary Figure~S4, the jump in flux density from non-detection to detection requires a temporal power law slightly steeper than $t^{4}$, which requires a CNM density profile with a power-law steeper than $k=2.8$. However, note that while this model predicts that the optically thick emission is rising, in fact, upon detection it is declining. The optically thin emission, on the other hand, is expected to be declining according to $F_{\nu}\propto t^{-3(p+2)/(2(5-k))}$ or $F_{\nu}\propto t^{(42-28k)/14}$ in either the relativistic jet or the sub relativistic outflow scenario, respectively. The steepest decline in this case for a $k=2.5$ is $t^{-3}$, while the observed emission decline rate becomes steeper than this.

We next turn to examine numerical emission models for off-axis relativistic jets. For that purpose, we use the publicly available\footnote{https://cosmo.nyu.edu/afterglowlibrary/boxfit2011.html} BoxFit code\cite{van_eerten_2012}. In this scenario, a steep rise in the radio emission (steeper than the $t^3$ above) can occur in large off-axis angles when the relativistic jet travels in a CNM with a constant density (a wind-like density results in a shallower rise\cite{granot_2018}). We therefore explore the constant ISM density scenario next. We find that the numerical model can provide a reasonable fit to individual epochs of the broadband spectra only. However, the best-fit parameters vary substantially between the fitted model of each epoch. For example, the optically thin spectrum observed on day $233$, requires an off-axis angle of $\sim 1.1$\,rad , while the radio emission observed on day $197$ requires an off-axis angle of $1.57$\,rad and and ISM density an order of magnitude higher than the one in the model for day $233$. Moreover, no numerical solution that can explain both the initial steep rise in flux density, and the spectral and temporal evolution, can be found.

In Supplementary Figure~S4, we show several temporal power-laws for the rise and the decline of the emission, including the above steepest rise and decline rates. 
As seen in the figure, assuming that a shockwave was launched at the time of optical discovery, leading to radio emission that peaks at the time we discovered the radio flare, then the radio emission should have been detectable at the time of the third null-detection observation. The observed decline rate of the radio emission after the discovery is also steeper than the one expected from standard theoretical models.

\section*{A Comparison Between the X-ray and Radio Emission Temporal Evolution}\label{sec:xray}

The X-ray luminosity of ASASSN-15oi [ref.\citen{Gezari_2017}] is shown in Figure~\ref{fig:xray} in units of the Eddington luminosity. We estimate the Eddington luminosity using the black hole mass estimate of $\sim 10^{6}\,{\rm M}_{\odot}$ [ref.\citen{Gezari_2017,holoien_2018}] corresponding to $L_{Edd} \sim 1.2\times 10^{44}\,{\rm erg\,s}^{-1}$. As seen in Figure~\ref{fig:xray}, there is a gap in the X-ray data, when we first detect the radio emission. Despite this, it seems that there is no apparent correlation between the X-ray and radio emissions. The X-ray emission is detected already shortly after the optical discovery of ASASSN-15oi, and slowly increased by a factor of $\sim 10$ up to about a year later and crosses the $1\%$ Eddington luminosity level.

In addition to previously published X-ray data\cite{Gezari_2017,holoien_2018}, short {\it Swift} snapshot observations were undertaken on 2019 and 2020 and are publicly available. The data were analyzed following the same procedures used to analyze past {\it Swift} observations of ASASSN-15oi [ref. \citen{Gezari_2017,holoien_2018}]. Due to limited sensitivity, we combined the data sets obtained over several months in 2019, resulting in a X-ray flux measurement of $8.07\pm1.1 \times 10^{-14}{\rm erg\,cm}^{-2}\,{\rm s}^{-1}$. The combined observations in 2020 show that the X-ray emission then faded to a flux level of $1.8\pm1.9 \times 10^{-14}{\rm erg\,cm}^{-2}\,{\rm s}^{-1}$.




\clearpage

\begin{figure*}
\centering
\includegraphics[width=\textwidth]{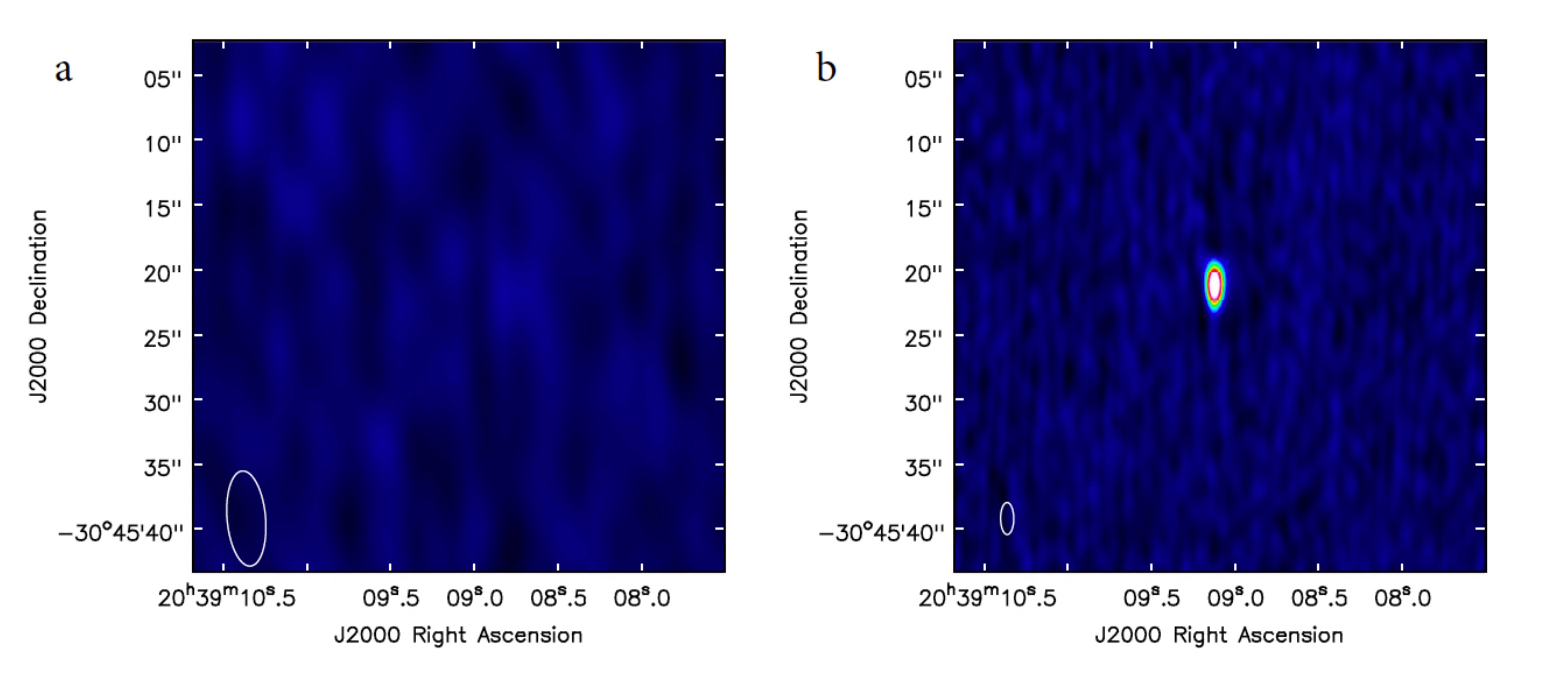}
\caption{\textbf{VLA K-band images of the position of the optical TDE candidate ASASSN-15oi, before and after radio detection}. The left panel (a) presents the third VLA image we obtained of this field $3$\,months after optical discovery on 2015 Nov 12, still showing a null-detection. The right panel (b) presents the image from our forth VLA observation on 2016 Feb 12 which reveals a delayed radio flare, $6$\,months after optical discovery. The synthesised beam size is shown as a white ellipse at the bottom left corner of the images. The flux density scale is identical in both images.}
\label{fig:radio_image}
\end{figure*}

\clearpage

\begin{figure*}
    \centering
        \includegraphics[width=1\linewidth]{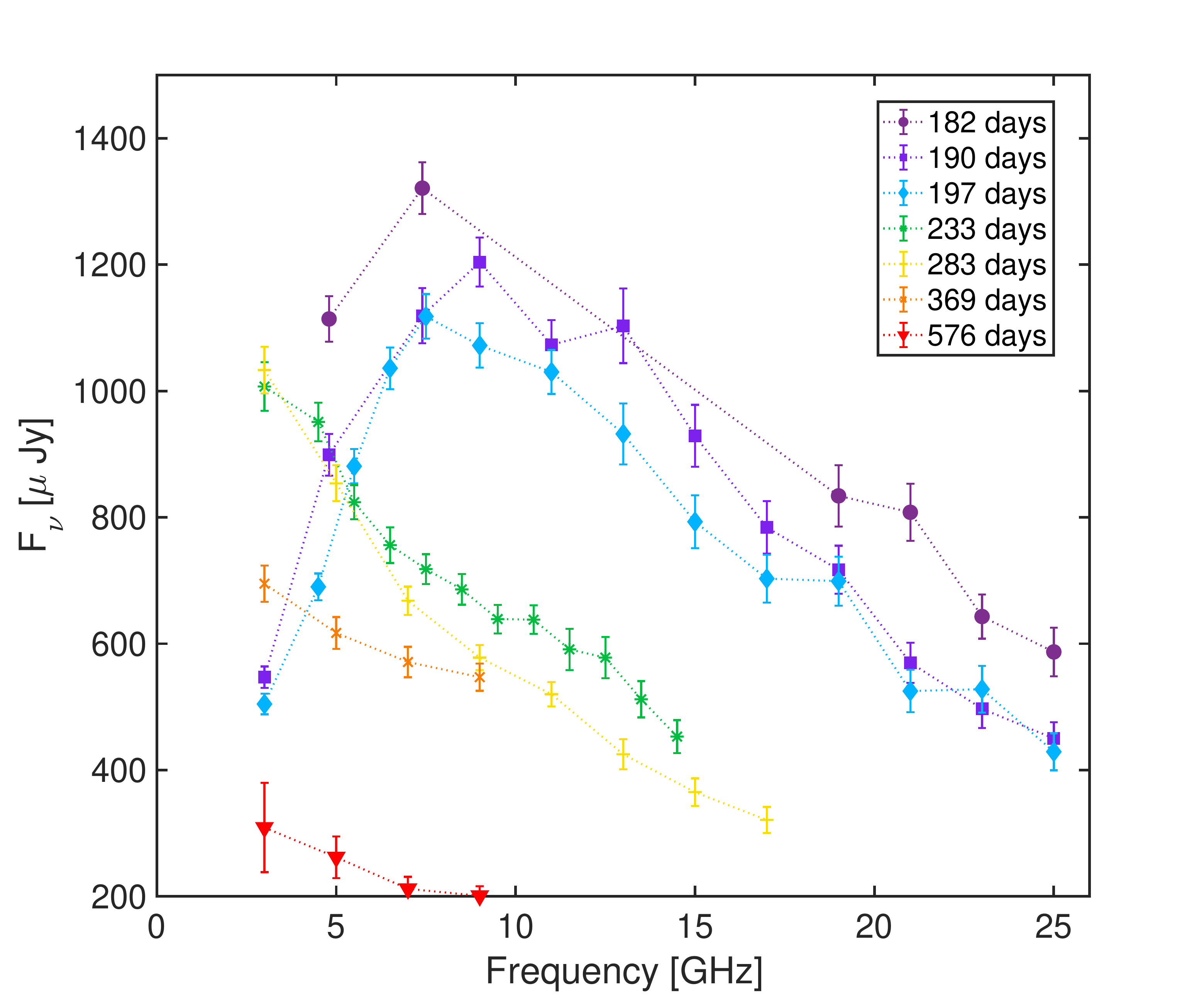}
    \caption{\textbf{The full observed broadband spectral evolution of the delayed radio flare from ASASSN-15oi.} Each of the radio broadband spectra is from a different observing epoch, starting from the initial detection of the delayed flare on $182$\,days and up to $576$\,days after optical discovery. Data from each epoch is represented by a different marker shape and color as noted in the legend (a dashed line connecting the data has been added for convenience). The error bars represent the image noise and flux calibration error added in quadrature (see Supplementary Table~1). 
    }
    \label{fig:radio_spec}
\end{figure*}

\clearpage

\begin{figure*}
    \centering
    \includegraphics[width=\linewidth]{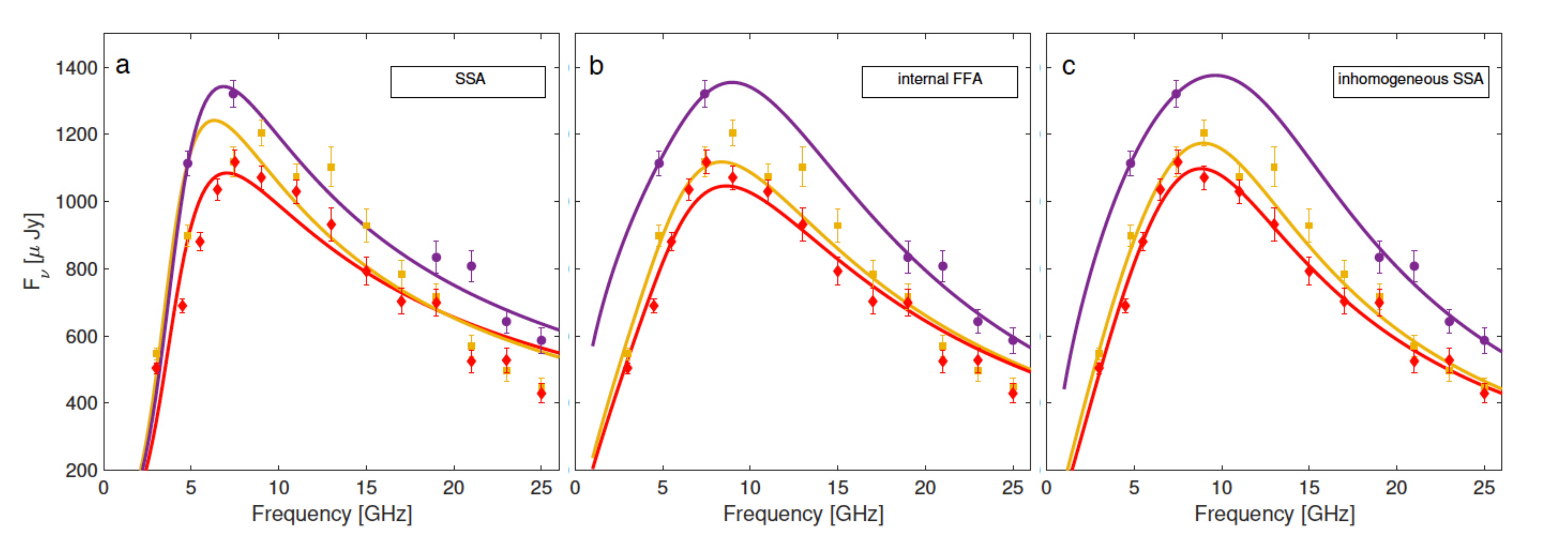}
\caption{\textbf{Best fit single-epoch spectral models of the radio flare spectra.} Observing epochs at $\Delta t=182, 190, 197$\,days are represented in purple, yellow and red, respectively. The broadband spectrum in each single epoch was fitted independently, thus not including any modeling of the temporal evolution. The errors of the data modeled here include the flux density calibration error and image noise added in quadrature. The left panel (a) presents the best fit homogeneous SSA model\cite{chevalier_1998}. The middle panel (b) shows the best fit models of the radio flare spectra using the internal free-free absorption model\cite{weiler_2002}. The right panel (c) is the best fit models using the inhomogeneous SSA model\cite{bjornsson_2017}. Out of the three models that we try here, the latter model is the best match to the  spectral data presented in this figure (see details in Methods).
}
    \label{fig:spec_fit}
\end{figure*}

\clearpage

\begin{figure*}
    \centering
    \includegraphics[width=1\linewidth]{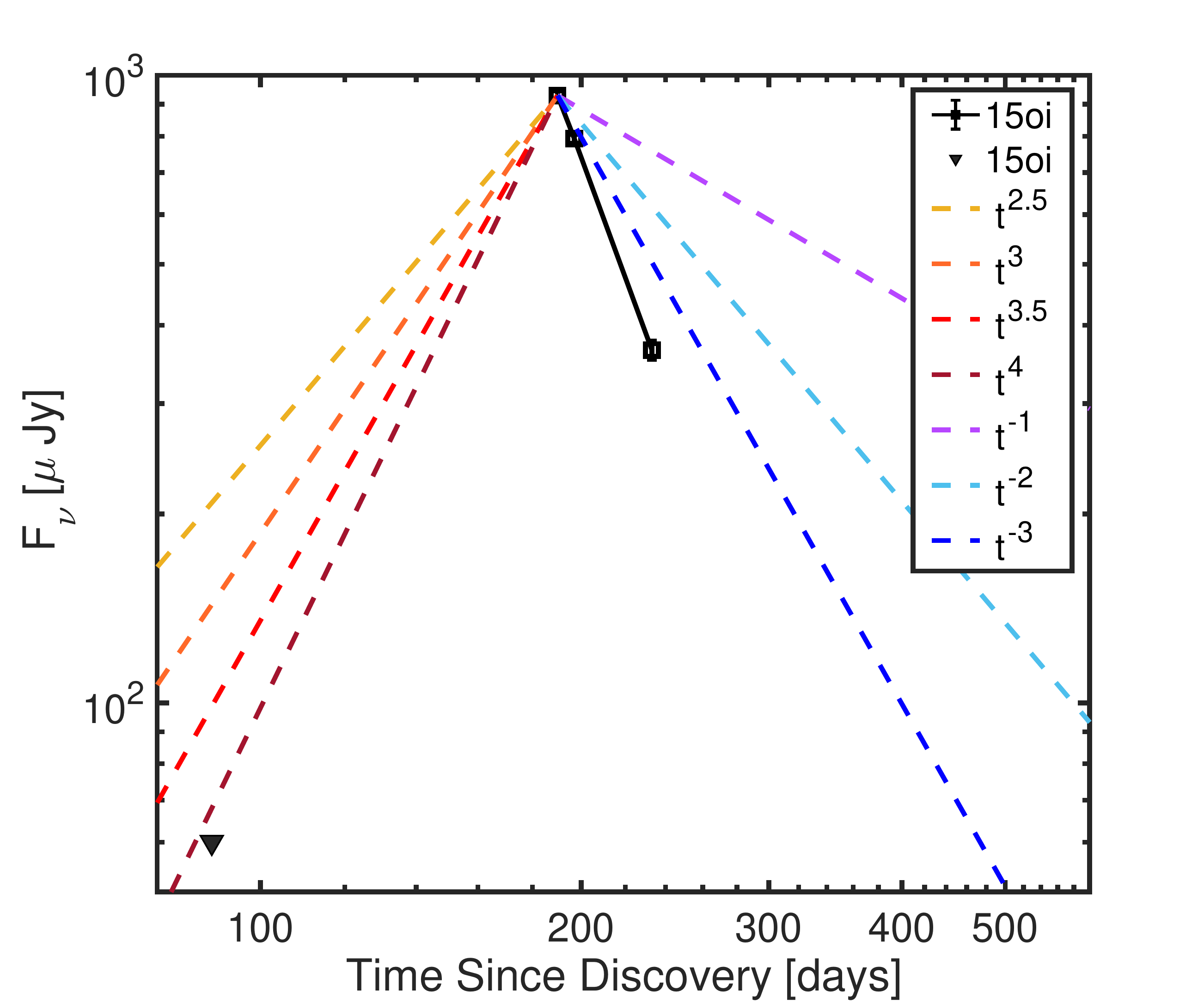}
     \caption{\textbf{Comparison of the the temporal evolution of the observed optically thin radio emission with different rising and declining power-law functions.} The presented radio emission is at a frequency of $15$\,GHz (black solid line and markers). The various power-law functions for both the rise of the emission (since the last non detection) and its decline are presented as dashed curves (representing various predictions, see details in Methods). The black triangle represents a 3$\sigma$ non-detection limit (based on the average between the $22$\,GHz and $6$\,GHz limits). 
     }
    \label{fig:afterglow_compare}
\end{figure*}

\clearpage


\clearpage

\begin{table*}
\small
    \renewcommand{\arraystretch}{0.5}

\centering
\caption{ASASSN-15oi VLA observation log}
\medskip
\begin{tabular}{lccc}
\hline
\smallskip
\\
Date & Frequency & Flux Density & Flux Density Error  \\
  & [GHz] & [$\mu$Jy] &  [$\mu$Jy]\\
\\
\hline
\\
2015 Aug 22 & 6.1 & $\leq 33$ & 11 \\
 & 22 & $\leq 60$ & 20 \\
2015 Sep 06 & 6.1 & $\leq 36$ & 12 \\
 & 22 & $\leq 69$ & 23 \\
2015 Nov 12 & 6.1 & $\leq 60$ & 20 \\
 & 22 & $\leq 33$ & 11 \\
2016 Feb 12 & 4.8 & 1114 & 13 \\
 & 7.4 & 1321 & 11 \\
 & 19 & 834 & 25 \\
 & 21 & 808 & 20 \\
 & 23 & 643 & 14 \\
 & 25 & 587 & 25 \\
2016 Feb 20 & 3 & 547 & 5 \\
 & 4.8 & 899 & 19 \\
 & 7.4 & 1119 & 28 \\
 & 9 & 1204 & 14 \\
 & 11 & 1073 & 22 \\
 & 13 & 1103 & 21 \\
 & 15 & 929 & 16 \\
 & 17 & 784 & 14 \\
 & 19 & 717 & 13 \\
 & 21 & 570 & 14 \\
 & 23 & 497 & 18 \\
 & 25 & 450 & 12 \\
 2016 Feb 27 & 3 & 504 & 6 \\
 & 4.5 & 690 & 5 \\
 & 5.5 & 881 & 8 \\
 & 6.5 & 1036 & 11 \\
 & 7.5 & 1118 & 11 \\
 & 9 & 1072 & 14 \\
 & 11 & 1030 & 16 \\
 & 13 & 932 & 12 \\
 & 15 & 793 & 13 \\
 & 17 & 703 & 14 \\
 & 19 & 717 & 14 \\
 & 21 & 699 & 17 \\
 & 23 & 525 & 21 \\
 & 25 & 429 & 20 \\
 2016 Apr 03 & 3 & 1007 & 24 \\
 & 4.5 & 951 & 11 \\
 & 5.5 & 824 & 11 \\
 & 6.5 & 756 & 17 \\
 & 7.5 & 718 & 10 \\
 & 8.5 & 686 & 13 \\
 & 9.5 & 639 & 12 \\
 & 10.5 & 638 & 12 \\
 & 11.5 & 591 & 14 \\
 & 12.5 & 578 & 15 \\
 & 13.5 & 512 & 13 \\
 & 14.5 & 453 & 13 \\
 2016 May 23 & 3 & 1033 & 20 \\
 & 5 & 854 & 12 \\
 & 7 & 668 &  10 \\
 & 9 & 578 & 10 \\
 & 11 & 520 & 11 \\
 & 13 & 425 & 11 \\
 & 15 & 365 & 12 \\
 & 17 & 321 & 13 \\
 2016 Aug 17 & 3 & 695 & 20 \\
 & 5 & 617 & 17 \\
 & 7 & 571 &  17 \\
 & 9 & 547 & 14 \\
 2017 Mar 12 & 3 & 309 & 70 \\
 & 5 & 262 & 32 \\
 & 7 & 212 &  18 \\
 & 9 & 201 & 14 \\
\\
\hline
\end{tabular}
      \small
      \begin{tablenotes}
\item The observation log includes both the initial null-detection $3\sigma$ limits and the following broadband observations in various bands. The flux density error quoted is the image rms noise. A flux density calibration error of $3\%$ and $5\%$ (below and above KU-band) should be added in quadrature.
 \end{tablenotes}
 
\end{table*}

\end{document}